\documentclass[trackchanges]{aastex63}
\hypersetup{linkcolor=red,citecolor=green,filecolor=cyan,urlcolor=magenta}

\received{}
\revised{}
\accepted{}
\submitjournal{PASP}

\shorttitle{Two-component model of $\gamma$-ray}
\shortauthors{Pei et al.}

\begin{document}

\title{Beamed and unbeamed emission of $\gamma$-ray blazars}

\correspondingauthor{Junhui Fan}
\email{fjh@gzhu.edu.cn}

\author[0000-0002-4970-3108]{Zhiyuan Pei}
\affiliation{Dipartimento di Fisica e Astronomia ``G.\ Galilei'', Universit\`a di Padova, I-35131 Padova, Italy}
\affiliation{Istituto Nazionale di Fisica Nucleare, Sezione di Padova, I-35131 Padova, Italy}
\affiliation{Center for Astrophysics, Guangzhou University, Guangzhou 510006, China}
\affiliation{Astronomy Science and Technology Research Laboratory of Department of Education of Guangdong Province, Guangzhou 510006, China}
\affiliation{Key Laboratory for Astronomical Observation and Technology of Guangzhou, Guangzhou 510006, China}

\author{Junhui Fan}
\affiliation{Center for Astrophysics, Guangzhou University, Guangzhou 510006, China}
\affiliation{Astronomy Science and Technology Research Laboratory of Department of Education of Guangdong Province, Guangzhou 510006, China}
\affiliation{Key Laboratory for Astronomical Observation and Technology of Guangzhou, Guangzhou 510006, China}

\author{Jianghe Yang}
\affiliation{Department of Physics and Electronics Science, Hunan University of Arts and Science, Changde 415000, China}

\author{Denis Bastieri}
\affiliation{Dipartimento di Fisica e Astronomia ``G.\ Galilei'', Universit\`a di Padova, I-35131 Padova, Italy}
\affiliation{Istituto Nazionale di Fisica Nucleare, Sezione di Padova, I-35131 Padova, Italy}
\affiliation{Center for Astrophysics, Guangzhou University, Guangzhou 510006, China}

\begin{abstract}
A two-component model of radio emission has been used to explain some radio observational properties of Active Galactic Nuclei (AGNs) and, in particular, of \emph{blazars}. In this work, we extend the two-component idea to the $\gamma$-ray emission and assume that the total $\gamma$-ray output of blazars consists of relativistically beamed and unbeamed components. The basic idea leverages the correlation between the radio core-dominance parameter and the $\gamma$-ray beaming factor. To do so, we evaluate this correlation for a large sample of 584 blazars taken from the fourth source catalog of the {\it Fermi} Large Area Telescope ({\it Fermi}-LAT) and correlated their $\gamma$-ray core-dominance parameters with radio core-dominance parameters. The $\gamma$-ray beaming factor is then used to estimate the beamed and unbeamed components. Our analysis confirms that the $\gamma$-ray emission in blazars is mainly from the beamed component.
\end{abstract}


\section{Introduction} 

Blazars, the most extreme subclass of AGNs, are characterized by large and rapid variability, apparent superluminal motion in their pc-scale jet, strong polarization, high energy $\gamma$-ray emission. All of these properties are affected by relativistic beaming. The emission in the jet is highly boosted along the observer's line of sight \citep{UP95}. Multiwavelength observational data show that the radio spectra of most blazars is usually flat with a power-law index $\alpha \simeq 0$ ($S_{\nu}\propto\nu^{-\alpha}$). According to the optical emission line features, blazars can be divided into flat spectrum radio quasars (FSRQs) and BL Lacertae objects (BL Lacs), where BL Lacs have weak or no emission lines (i.e. the equivalent width, EW, of the emission line in rest frame is less than 5 $\mathring{A}$), while FSRQs show stronger emission lines \citep[EW $\ge5\:\mathrm{\AA}$,][]{Sto91, Sti91, UP95} in their optical spectra. From the transition between FSRQs and BL Lacs, and based on the luminosity of the broad-line region (BLR) measured in Eddington units, \citet{Ghi11} proposed a physical distinction between these classes, setting the dividing line of $L_{\rm BLR}/L_{\rm Edd}\sim5\times10^{-4}$. 

Based on a relativistic beaming model, \citet{US84} proposed that the total emission from AGNs consists of two components, namely, a beamed component and an unbeamed one. Then, the observed total luminosity, $L^{\rm tot}$, is the sum of the beamed, $L_{\rm b}$, and unbeamed, $L_{\rm unb}$ contributions, i.e.\ $L^{\rm tot}=L_{\rm b}+L_{\rm unb}$. In the radio band, the ratio of the two components, $R_{\rm radio}$, is defined as the radio core-dominance parameter, i.e.\ $R_{\rm radio}=L_{\rm b}/L_{\rm unb}$ \citep[see][and reference therein]{OB82, Fan11, Pei16, Pei19, Pei20(b), Pei20(a)}. This ratio can also be expressed as:
\begin{equation}
R(\phi) = \displaystyle \frac{\textrm{flux density of beamed compact core}}{\textrm{flux density of unbeamed components}},\,\,\,\,\,\,R_\perp = R(90^{\circ}),
\label{eq1}
\end{equation} 
where $\phi$ is the viewing angle between the line-of-sight and the motion direction of the approaching side of the compact core and $R_\perp$ refers to the ratio of the luminosity in the jet to the unbeamed luminosity when $\phi = 90^{\circ}$. The compact core is assumed to be produced by emission from the unresolved bases of two oppositely directed jets moving with speed $\beta c$ relative to the central object, where $\beta$ is the relativistic bulk velocity in units of the speed of light $c$ \citep{OB82, US84, UP95}.

Previous studies have shown that the radio core-dominance parameter $R_{\rm radio}$ can play the role of a tracer of the beaming effect, $R_{\rm radio}=f\delta^{p}$ \citep[see][]{UP95, Fan03}, where $f$ is the intrinsic ratio, defined by the intrinsic flux density of the jet to the extended flux density in the co-moving frame, $f= \frac{S^{\rm in}_{\rm core}}{S^{\rm in}_{\rm ext.}}$. $\delta$ is the Doppler factor, $p=n+\alpha$, $\alpha$ is the radio spectral index and $n$ depends on the shape of the emitted spectrum and the physical details of the jet, being $n = 2$ for a continuous jet and $n = 3$ for a moving blob. The Doppler factor $\delta$ is $\delta=[\Gamma(1-\beta\cos\phi)]^{-1}$, where $\Gamma$ is the Lorentz factor defined as $\Gamma=(1-\beta^{2})^{-1/2}$. Then, the radio core-dominance parameter can be written as $R_{\rm radio}=f\delta^{p}=f[\Gamma(1-\beta\cos\phi)]^{-p}$. When the viewing angle $\phi$ is large ($\phi\gtrsim\arccos[0.5/\beta]$, see \citet{UP95}), the emission from the receding jet is no longer negligible, and the relation becomes \citep{UP95, Fan03}  
\begin{equation}
R_{\rm radio}=f\Gamma^{-p}[(1-\beta\cos\phi)^{-p} + (1+\beta\cos\phi)^{-p}].
\label{eq2}
\end{equation}

The Doppler factor is a crucial parameter in blazars, since it indicates how strongly the flux density is boosted and the variability timescales shortened in the observer's frame. However, it is difficult to determine this parameter because it is not possible to measure it directly. Based on $R_{\rm radio}=f\delta^{p}$ or relation (\ref{eq2}), we can take the radio core-dominance parameter $R_{\rm radio}$ as a statistical indicator for studying the relativistic beaming effect if the ratio $f$ is known.

In 1987, \citet{BM87} proposed that the observed X-ray emission should be also formed of two components, the beamed and unbeamed ones. These authors assumed that the beamed X-ray radiation originates in the compact radio source, while the total X-ray emission is proportional to the extended radio emission. \citet{K93} proposed a different model to separate the total X-ray luminosity into beamed and isotropic components and obtained a tight correlation between the total X-ray luminosity and the radio extended one by studying 34 radio loud quasars from \citet{BM87}. He found that the beamed X-ray emission is associated with the radio emission and obtained $R_{X\perp}=1.30\times10^{-2}$. \citet{Fan05} applied the two-component model of X-ray emission to 19 selected $\gamma$-ray loud blazars and showed that the beamed emission dominates the X-ray flux, while the unbeamed X-ray emission is correlated with the extended radio emission, and they ascertained $R_{X\perp}=5.90\times10^{-3}$.      

After the launch of {\it Fermi} Large Area Telescope (hereafter, {\it Fermi}-LAT), many new high-energy $\gamma$-ray sources were detected, increasing our knowledge about $\gamma$-ray blazars and opening new opportunities to study the $\gamma$-ray production mechanisms. In particular, the {\it Fermi}-LAT collaboration released the 4FGL catalog, based on the first eight years of data taking. This catalog includes 5098 sources above the significance of $4\sigma$, covering the 50 MeV$-$1 TeV range \citep{4FGL, 4LAC}.  AGNs are the vast majority of sources in the 4FGL; among them 2938 are blazars: 681 FSRQs, 1102 BL Lacs and 1152 blazar candidates of unknown class (BCUs)\citep{4FGL}. 

A widespread idea is that the $\gamma$-ray emission of powerful blazars is mainly produced within the broad-line region (BLR) from external-inverse Compton interaction \citep[EC, e.g.,][]{Sik94, GM96, Baz00, Der09, Pal15} while in low-power blazars it is due to the synchrotron self-Compton mechanism \citep[SSC, e.g.,][]{Mar92, Tav98, Fin08}. 

\citet{WuD14} found no correlation between the $\gamma$-ray luminosity and the radio core-dominance parameter in a sample of 124 $\gamma$-ray blazars, however, they found that the ratio of $\gamma$-ray luminosity to the extended radio luminosity, $\log (L_{\gamma}/L_{\rm ext})$, is correlated with the radio core-dominance parameter, with a positive linear regression given by $\log (L_{\gamma}/L_{\rm ext}) \sim 0.946 \log (1+R_{\rm radio})$ and proposed that the $\gamma$-ray luminosity consists of two components, similarly to the behaviour in the radio and X-ray bands.

Later, \citet{Pei16} compiled a sample of 169 $\gamma$-ray loud blazars from 3FGL \citep{3FGL} with available radio core-dominance parameters ($R_{\rm radio}$) and $\gamma$-ray photon indices ($\alpha^{\rm ph}_{\gamma}$), and investigated a two-component model for the $\gamma$-ray emission. \citet{Pei16} obtained a correlation between $\alpha^{\rm ph}_{\gamma}$ and $R_{\rm radio}$, as done for the radio band by \citet{Fan10},
\begin{equation}
\alpha^{\rm ph}_{\gamma, \rm total}=\displaystyle\frac{R_{\rm radio}}{1+R_{\rm radio}}\alpha^{\rm ph}_{\gamma, \rm core}+\frac{1}{1+R_{\rm radio}}\alpha^{\rm ph}_{\gamma, \rm ext}.
\label{eq3}
\end{equation}
Here $\alpha^{\rm ph}_{\gamma, \rm total}$ denotes the $\gamma$-ray photon index listed in the 3FGL catalog, $\alpha^{\rm ph}_{\gamma, \rm core}$ and $\alpha^{\rm ph}_{\gamma, \rm ext}$ denote the $\gamma$-ray photon indices of the core and the extended components, respectively. Assuming that all of the sources follow equation (\ref{eq3}), then one can estimate $\alpha^{\rm ph}_{\gamma, \rm core}$ and $\alpha^{\rm ph}_{\gamma, \rm ext}$ for BL Lacs and FSRQs by minimizing $\Sigma\left[\alpha^{\rm ph}_{\gamma, \rm total}-\alpha^{\rm ph}_{\gamma, \rm core}R_{\rm radio}/(1+R_{\rm radio})-\alpha^{\rm ph}_{\gamma, \rm ext}/(1+R_{\rm radio})\right]^{2}$ \citep{Fan10}. When equation (\ref{eq3}) was applied to the sample, \citet{Pei16} obtained the fitting results that $\alpha^{\rm ph}_{\gamma, \rm core} = 2.09\pm0.01$ and $\alpha^{\rm ph}_{\gamma, \rm ext} = 1.58\pm0.06$ for BL Lacs. On the other hand, for FSRQs, $\alpha^{\rm ph}_{\gamma, \rm core} = 2.34\pm0.01$ and $\alpha^{\rm ph}_{\gamma, \rm ext} = 1.72\pm0.01$. The average values of $\alpha^{\rm ph}_{\gamma, \rm total}$ for the sub-sample of BL Lacs were $2.10\pm0.21$ and $2.38\pm0.21$ for FSRQs, which were obtained from the 3FGL. Comparing with the numerical fitting results and the average values, it was found that the $\gamma$-ray photon indices derived for the core components are approximately equal to the statistical average values of the photon indices for BL Lacs and FSRQs, respectively. These results supported that the $\gamma$-ray emission is mainly from the core component when a two-component model is considered for the the $\gamma$-ray band.

It is quite interesting to note that the nearest TeV-detected radio galaxy, Centaurus A, was first reported by the High Energy Stereoscopic System (H.E.S.S.) at TeV energies \citep{HESS09}, and only later by {\it Fermi}-LAT at the GeV band \citep{Fer10}. Fanaroff-Riley type I (FRI) radio galaxies are the parent population of BL Lacs, and Fanaroff-Riley type II (FRII) radio galaxies are the parent population of FSRQs \citep{UP95}. {\it Fermi} also discovered the spatial extension within Cen A (FRI radio galaxy) by detecting the $\gamma$-ray emission both from the core and extended giant lobes, which implies that the $\gamma$-rays are from both the central core and extended kilo-parsec scale lobes \citep{Fer10}. It suggests that the $\gamma$-rays in BL Lacs are both from the core and from the extended lobes (two components). We can also assume that the $\gamma$-ray emission in FSRQs is the superposition of the core and the extended components since FRII radio galaxies and FSRQs come from the same population. Therefore, it is reasonable to assume that the $\gamma$-rays are also from two different components.

In this paper, we propose the hypothesis that the $\gamma$-ray emission consists of two components (the beamed and unbeamed ones) and separate them in much the same way as it has been done for the radio and X-ray radiation. The methodology used is discussed in Sect.\ \ref{sec2}, while in Sect.\ \ref{sec3} we describe the sample and the results. In Sect. \ref{sec4} we present the statistical analysis and discuss the results, while drawing the conclusions in Sect.\ \ref{sec5}. In this paper, we will apply the $\Lambda$CDM model, with $\Omega_{\Lambda} \simeq 0.73$, $\Omega_{M} \simeq 0.27$, and $H_{0} \simeq 73 \rm {km \cdot s^{-1} \cdot Mpc^{-1}}$.

\section{METHODOLOGY}\label{sec2}

\subsection{$\gamma$-ray core-dominance parameter}
\citet{OB82} found a relation among parameters $R_\perp$, $f$, and $\Gamma$ given by:
\begin{equation}
R_\perp=\displaystyle\frac{2f}{\Gamma^{p}}.
\label{eq4}
\end{equation}
After substituting for parameter $f$ and considering flat-spectrum radio AGNs (i.e.\ $\alpha=0$, resulting in $p=2$), the relation (\ref{eq2}) can be expressed as:
\begin{equation}
R_{\rm radio} = R_\perp\displaystyle\frac{1}{2}\left[(1 - \beta \cos \phi)^{-2} + (1 + \beta \cos \phi)^{-2}\right],
\label{eq5}
\end{equation}
where the second term on the right-hand side originates from a ``counter-jet''. If one defines $g(\beta, \phi)=\displaystyle\frac{1}{2}\left[(1 - \beta \cos \phi)^{-2+\alpha} + (1 + \beta \cos \phi)^{-2+\alpha}\right]$ \citep{BM87}, then equation (\ref{eq5}) can be reduced to $R=R_\perp g(\beta, \phi)$ ($\alpha=0$). For a source, if the values of $\beta$, $\phi$ and $R_{\rm radio}$ are given, one can obtain $R_\perp$ from equation (\ref{eq5}). Using this idea, \citet{OB82} used a sample of 32 quasar to obtain an estimate of $R_\perp=0.024$. Therefore they concluded that $R_\perp$ should be taken as a constant with a value of 0.024. \citet{K93} and \citet{Fan05} also adopted $R_\perp=0.024$ to probe the beamed and unbeamed components of X-rays emission. In this work, we take the value $R_\perp=0.024$ as concluded by \citet{OB82} and accepted by \citet{K93} and \citet{Fan05}. Adopting the similar method as shown in \citet{OB82}, \citet{Fan03} also obtained $R_\perp = 0.044$ by studying 38 superluminal sources and concluded that $R_\perp < 0.1$ was satisfied by most objects.

We adopt the two-component model used for the radio and X-ray emission \citep[see][]{BM87, K93, Fan05}. In this scenario, the ratio of the beamed radio emission in the transverse direction to the extended radio  emission is considered to be constant \citep{OB82}. Thus it is expected that the transverse beamed $\gamma$-ray luminosity, $L_{\gamma,\rm b\perp}$ is proportional to the extended radio emission, $L_{\rm r,unb}$, i.e. $L_{\gamma,\rm b\perp} = A \cdot L_{\rm r,unb}$, where $A$ is a constant. Similar to the radio beaming factor, we can define a parameter, namely the $\gamma$-ray beaming factor $g_{\gamma}(\beta, \phi)$, as 
\begin{equation}
g_{\gamma}(\beta, \phi) = \frac{1}{2}\left[(1 - \beta \cos \phi)^{-(2+\alpha_{\gamma})} + (1 + \beta \cos \phi)^{-(2+\alpha_{\gamma})}\right],
\label{eq6}
\end{equation}
where $\alpha_{\gamma}$ is the $\gamma$-ray spectral index of the beamed emission ($\alpha_{\gamma} = \alpha_{\gamma}^{\rm ph}-1$). Therefore, the beamed luminosity for an inclination angle $\phi$ between the jet direction and the line of sight is $L_{\gamma,\rm b}(\phi) = g_{\gamma}(\beta, \phi) L_{\gamma,\rm b}(90^\circ)$. We now define the ratio of the beamed to unbeamed $\gamma$-ray luminosity, the $\gamma$-ray core-dominance parameter, as:
\begin{equation}
R_{\gamma} = \displaystyle \frac{L_{\gamma,\rm b}}{L_{\gamma, \rm unb}} = R_{\gamma\perp} g_{\gamma}(\beta, \phi), \,\, R_{\gamma\perp} = \displaystyle \frac{L_{\gamma,\rm b}(90^\circ)}{L_{\gamma,\rm unb}},
\label{eq7}
\end{equation}  
where $R_{\gamma\perp}$ is the ratio of the beamed to the unbeamed $\gamma$-ray component. We proceed in analogy with the beaming model of \citet{OB82} and \citet{K93} \citep[see also][]{Fan05} assuming that this parameter is constant. 

\begin{figure}
   \centering
   \includegraphics[width=8cm]{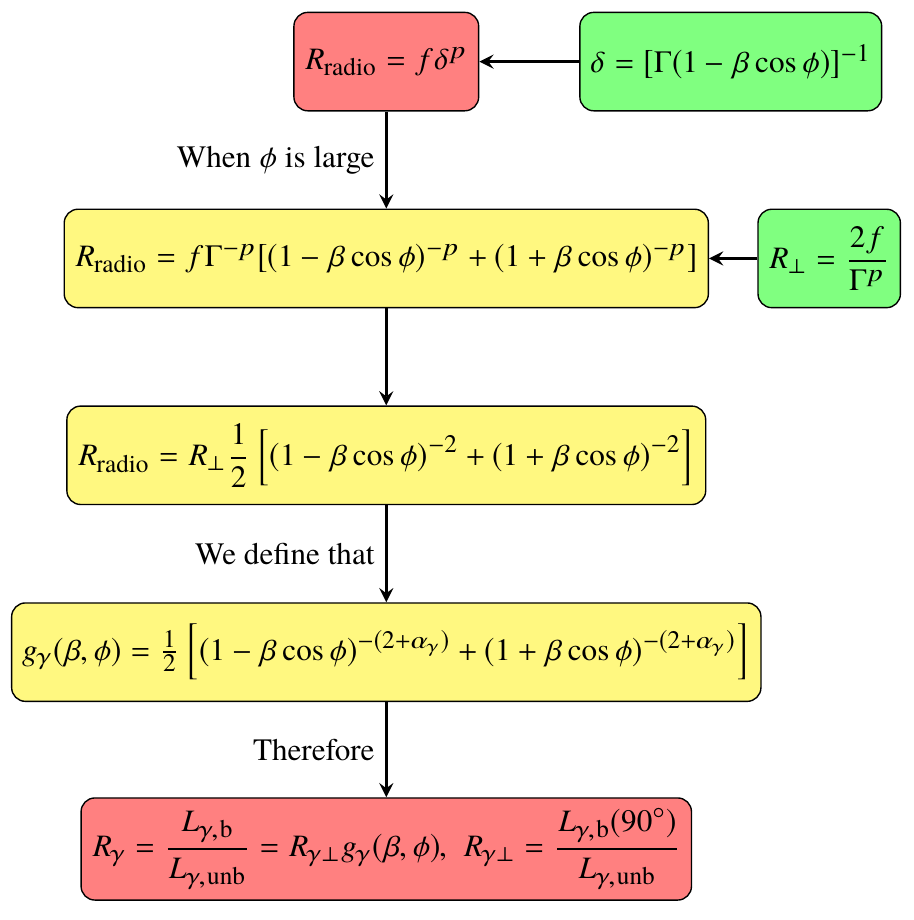}
   \caption{Flowchart of the scheme of calculation to derive the $\gamma$-ray core-dominance parameter ($R_{\gamma}$).}
     \label{fig1}
\end{figure}

For a given $\gamma$-ray blazar we can obtain $\beta\cos\phi$ from equation (\ref{eq5}): $\beta\cos\phi=\bigg(\frac{1}{2}\big\{2R_{\rm radio}+R_\perp-\big[R_\perp\left(8R_{\rm radio}+R_\perp\right)^{1/2}\big]\big\}\bigg)^{1/2}$, then, we can estimate the $\gamma$-ray beaming factor $g_{\gamma}(\beta, \phi)$ via equation (\ref{eq6}) when the $\gamma$-ray spectral index $\alpha_{\gamma}$ is available, and finally calculate the $\gamma$-ray core-dominance parameter $R_{\gamma}$ using equation (\ref{eq7}) following the flowchart shown in Fig.\ \ref{fig1}. 

\subsection{$\gamma$-ray luminosity} 

The $\gamma$-ray luminosity can be calculated from the detected photons and the distance to the source \citep{Abd10b, Fan13}. The $\gamma$-ray photon flux is assumed to follow a power-law spectrum \citep[see][]{Ghi93, Fan00, 3FGL, Sin15, 4FGL}, which can be expressed as:
\begin{equation}
\displaystyle \frac{dN}{dE} = N_{0} E^{-\alpha^{\rm ph}_{\gamma}},
\label{eq8}
\end{equation}
where $\alpha^{\rm ph}_{\gamma}$ is the $\gamma$-ray photon spectral index. Then, we can calculate the $\gamma$-ray luminosity by:  
\begin{equation}
L_{\gamma} = 4\pi d_{\rm L}^{2}(1+z)^{\alpha^{\rm ph}_{\gamma}-2}F,
\label{eq9}
\end{equation} 
where $d_{\rm L}$ is the luminosity distance, $(1+z)^{\alpha_{\gamma}^{\rm ph}-2}$ is for $K$-correction. The integral flux $F$ in units of $\mathrm{GeV\,cm^{-2}s^{-1}}$ can be obtained by $F=\int_{E_L}^{{E_U}}EdN$, where we adopt $E_{L}=1\:\mathrm{GeV}$ and $E_{U}=100\:\mathrm{GeV}$ in our calculation.

\section{SAMPLE AND RESULTS} \label{sec3}

\citet{Pei20(a)} analysed the radio core-dominance parameter, $\log R_{\rm radio}$, for a sample of 4388 AGNs, out of which 584 are {\it Fermi}-LAT-detected blazars from 4FGL \citep{4FGL}. \citet{Pei20(a)} listed available data about the core radio luminosity and the extended radio luminosity at 5 GHz for these 584 sources. The unresolved core component of radio luminosity is assumed to be the beamed component, while the extended component should be the unbeamed component \citep{OB82, US84, UP95}. In this paper, we consider the core radio luminosity $L_{\rm r,b}$ in units of $\mathrm{W\,Hz^{-1}}$ as the beamed radio emission contribution, and the extended radio luminosity $L_{\rm r,unb}$ also in units of $\mathrm{W\,Hz^{-1}}$ as the unbeamed radio emission contribution. Thus we have 584 {\it Fermi}-detected blazars with available both beamed and unbeamed radio emission contributions. We use these data to calculate the $\gamma$-ray core-dominance parameter $R_{\gamma}$. Using the data from the 4FGL catalog, the $\gamma$-ray luminosity, $L_{\gamma}$ in units of $\mathrm{W\,Hz^{-1}}$, can be calculated via equation (\ref{eq9}).   

\begin{figure}
   \centering
   \includegraphics[width=10cm]{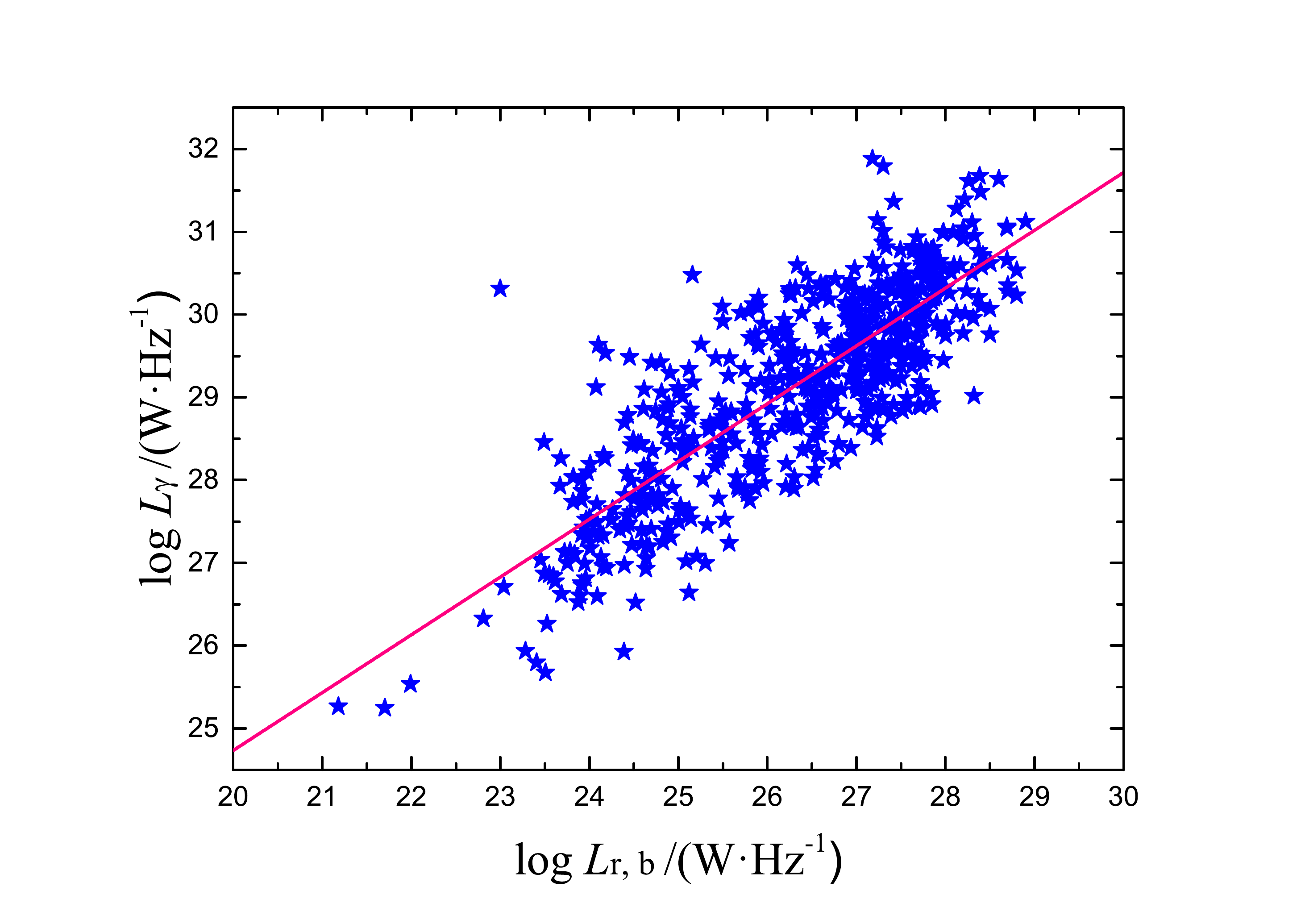}
   \caption{Plot of the $\gamma$-ray luminosity, $\log L_{\gamma}$, versus the core radio emission (beamed component) for the whole sample. The best-fitting gives $\log L_{\gamma} = (0.70 \pm 0.02)\log L_{\rm r,b} + (10.77 \pm 0.53)$.}
     \label{fig2}
\end{figure}

We applied a linear regression analysis to the luminosities for the 584 {\it Fermi} blazars and obtained
$$\log L_{\gamma} = (0.70 \pm 0.02)\log L_{\rm r, b} + (10.77 \pm 0.53),$$
with a correlation coefficient $r = 0.82$ and a chance probability $p \sim 0$. The result is shown in Fig.\ \ref{fig2}. From this regression we have $L_{\gamma} \propto L_{\rm r,b}^{0.70}$, indicating that the $\gamma$-ray emission is closely correlated with the relativistically beamed radio emission. Following \citet{K93} and \citet{Fan05}, we assumed a dominance of beamed emission in the $\gamma$-ray band \citep[e.g.,][]{Ghi93}, $L_{\gamma, \rm b} \simeq L_{\gamma} \propto L_{\rm r, b}^{0.70}$, valid for all $\gamma$-ray blazars, so
$$\log L_{\gamma, \rm b} = 0.70 \log L_{\rm r, b} + \log k,$$
where $k$ is a constant. Then the total $\gamma$-ray luminosity can be expressed as done by \citet{K93} for X-ray luminosity as
\begin{equation}
L_{\gamma} = L_{\gamma, \rm b} + L_{\gamma, \rm unb} = L_{\gamma, \rm b}\left(1+\displaystyle \frac{1}{R_{\gamma}}\right) = k L^{0.70}_{r, \rm b}\left[1+\displaystyle \frac{1}{R_{\gamma\perp} g_{\gamma}(\beta, \phi)}\right],  
\label{eq10}
\end{equation} 
where $L_{\gamma,\rm unb} = \displaystyle \frac{L_{\gamma,\rm b}}{R_{\gamma}} =\displaystyle \frac{L_{\gamma,\rm b}}{R_{\gamma\perp} g_{\gamma}(\beta, \phi)}$. The constants $k$ and $R_{\gamma\perp}$ can be determined by minimizing $\Sigma[\log(L_{\gamma}/L^{\rm obs}_{\gamma})]^{2}$ for all the sources \cite[see][]{BM87, K93, Fan05}, where $L^{\rm obs}_{\gamma}$ is the observed luminosity. For 584 {\it Fermi}-LAT blazars, we obtained the values with $\log k = 10.73$ and $R_{\gamma\perp} = 3.05 \times 10^{-3}$. Then, given the radio core-dominance parameter, $R_{\rm radio}$, we can obtain $\beta \cos \phi$ from equation (\ref{eq5}) using $R_\perp=0.024$ \citep{BM87, K93, Fan05}. This in turn gives the $\gamma$-ray beaming factor $g_{\gamma}(\beta, \phi)$ via equation (\ref{eq6}). Finally we can calculate the $\gamma$-ray core-dominance parameter $R_{\gamma}$ using equation (\ref{eq7}). Results are listed in Table \ref{tab1}, where column (1) gives the 4FGL name, column (2) the IAU name, column (3) the classification (BL: {\it Fermi} BL Lacerate objects; FSRQ: {\it Fermi} flat spectrum radio quasars and BCU: blazar candidates of unknown class), column (4) redshift, column (5) the $\gamma$-ray photon index with 1$\sigma$ uncertainty, column (6) the $\gamma$-ray luminosity with 1$\sigma$ uncertainty (in units of W Hz$^{-1}$), column (7) the radio core-dominance parameter, column (8) the reference for column (7) (P20: \citet{Pei20(a)}) and column (9) the derived $\gamma$-ray core-dominance parameter in this work. The table is available in its entirety in machine-readable form.

\begin{table}
\caption{\label{tab1} Sample of {\it Fermi} blazars}
\centering
\setlength{\tabcolsep}{3.5mm}{
\begin{tabular}{ccccccccc}
\hline\hline
4FGL Name & IAU Name & Class & $z$ & $\alpha^{\rm ph}_{\gamma}$ & $\log L_{\gamma}$ & $\log R_{\rm radio}$ & Ref. & $\log R_{\gamma}$\\
&&&&&(W Hz$^{-1}$)&&&\\
(1) & (2) & (3) & (4) & (5) & (6) & (7) & (8) & (9) \\
\hline

4FGL J0050.7$-$0929 	&	0048$-$097	&	BL	&	0.634	&	2.04/0.02 	&	29.94/0.01 	&	 1.20	&	P20 & 3.49	\\
4FGL J0108.6+0134 	&	0106+013	&	FSRQ	&	2.099	&	2.35/0.01 	&	31.67/0.01 	&	0.71 &	P20	&	2.91	\\
4FGL J0112.1+2245 	&	0109+224	&	BL	&	0.265	&	2.07/0.07 	&	 29.34/0.01	&	1.93 &	P20	&	5.02	\\
4FGL J0113.4+4948 	&	0110+495	&	FSRQ	&	0.389	&	2.23/0.05 	&	28.79/0.01 	&	 0.98 &	P20	&	3.32	\\
4FGL J0113.7+0225 	&	0111+021	&	BL	&	0.047	&	2.51/0.17 	&	25.92/0.03 &	 1.05 &	P20 	&	3.89	\\
4FGL J0114.8+1326 	&	0111+131	&	BL	&	0.685	&	2.07/0.04 	&	 29.50/0.01	& 2.00 &	P20	&	5.16	\\
4FGL J0116.0$-$1136 	&	0113$-$118	&	FSRQ	&	0.670 	&	2.39/0.04 	&	29.53/0.02 	&	1.02 &	P20 	&	3.64	\\
4FGL J0120.4$-$2701 	&	0118$-$272	&	BL	&	0.559 	&	1.90/0.02 	&	29.84/0.01 	& 1.11 &	P20	&	3.10	\\
$\cdots$ & $\cdots$ & $\cdots$ & $\cdots$ & $\cdots$ & $\cdots$ & $\cdots$ & $\cdots$ & $\cdots$ \\

\hline
\end{tabular}
}
\tablecomments{Column (1) gives the 4FGL name, column (2) the IAU name, column (3) the classification (BL: {\it Fermi} BL Lacerate objects; FSRQ: {\it Fermi} flat spectrum radio quasars and BCU: blazar candidates of unknown class), column (4) redshift, column (5) the $\gamma$-ray photon index with 1$\sigma$ uncertainty, column (6) the $\gamma$-ray luminosity with 1$\sigma$ uncertainty (in units of W Hz$^{-1}$), column (7) the radio core-dominance parameter, column (8) the reference for column (7) (P20: \citet{Pei20(a)}) and column (9) the derived $\gamma$-ray core-dominance parameter in this work.   
\\
(The table is available in its entirety in machine-readable form)\footnote{Readers can also send your request to {\it zhiyuan.pei@phd.unipd.it} for asking the full table.}}
\end{table}

\begin{figure}
   \centering
   \includegraphics[width=10cm]{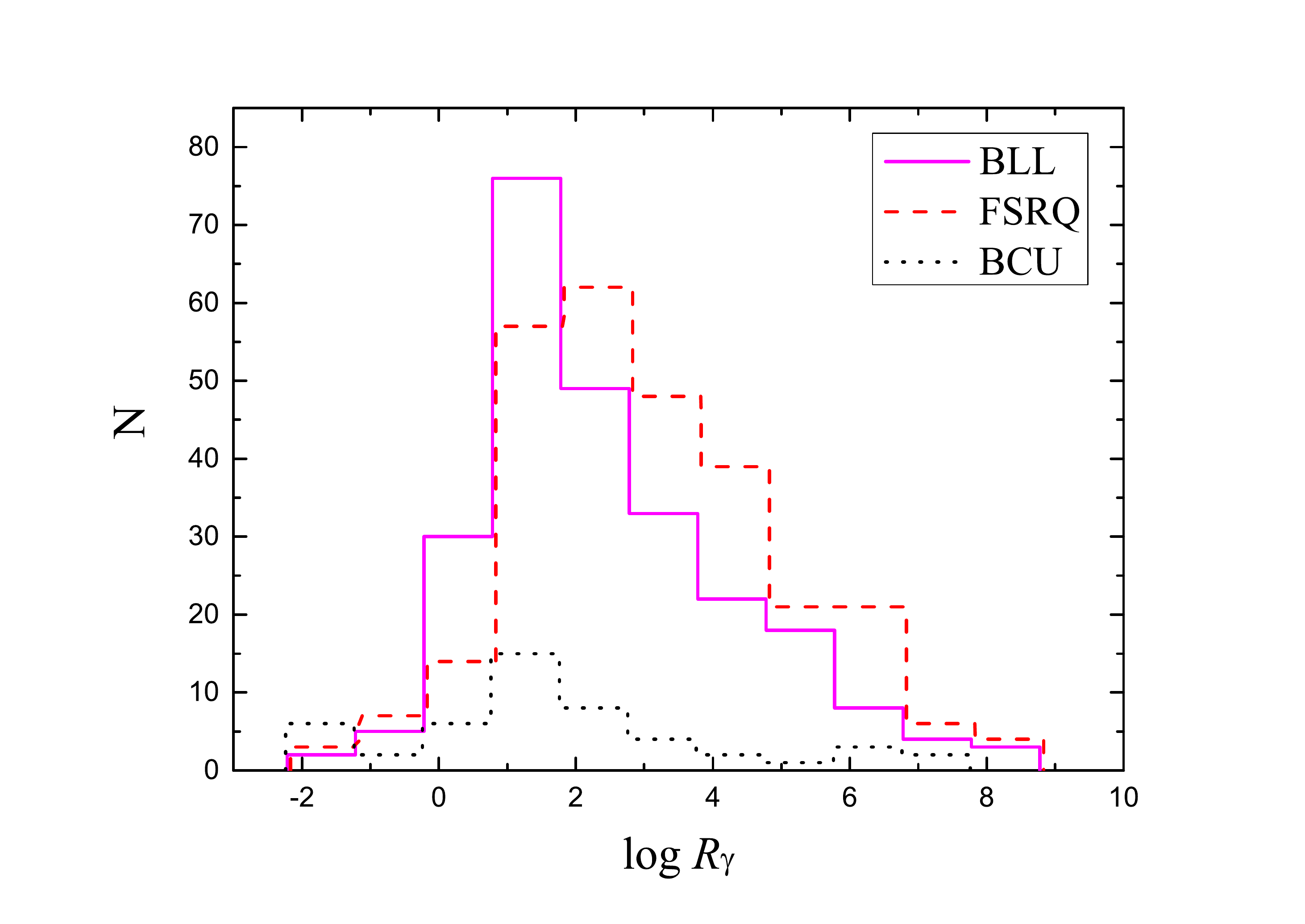}
   \caption{Distributions of $R_{\gamma}$ for BL Lacs, FSRQs and BCUs in our sample.}
     \label{fig3}
\end{figure}

The distribution of the $\gamma$-ray core-dominance parameter, $\log R_{\gamma}$, for BL Lacs, FSRQs and BCUs is shown in Fig. \ref{fig3}. We found that the average values of $\log R_{\gamma}$ for BL Lacs, FSRQs and BCUs are $\langle \log R_{\gamma} \rangle|_{\mathrm{BL\,Lac}} = 2.38 \pm 1.91$; $\langle \log R_{\gamma} \rangle|_{\rm FSRQ} = 3.07 \pm 2.01$ and $\langle \log R_{\gamma} \rangle|_{\rm BCU} = 1.80 \pm 2.42$, respectively. A Kolmogorov-Smirnov test (hereafter, K-S test) between the distributions of $\log R_{\gamma}$ for BL Lacs and FSRQs shows that they likely belong to different parent distributions ($p < 10^{-4}$). From the distributions and the K-S test result, we found that $\langle \log R_{\gamma} \rangle|_{\rm FSRQ}> \langle \log R_{\gamma} \rangle|_{\mathrm{BL\,Lac}}$, indicating that the $\gamma$-ray emission of {\it Fermi}-detected FSRQs are more core dominated than are {\it Fermi} BL Lacs.

\section{DISCUSSION} \label{sec4}

Blazars show extreme observational properties, which are associated with the relativistic beaming effect. Many methods are proposed to estimate the beaming boosting factor \citep[e.g.,][]{Ghi93, Xie02, Xie05, Fan99, Lio18, Xia19, Zha20, Pei20(c)}. In particular, the radio core-dominance parameter, $R_{\rm radio}$, has been found to be correlated with the polarization \citep{Wil92, Fan06}. 

The methodology of separating the luminosity into relativistically beamed and unbeamed components was successfully applied to the radio and X-ray emission of blazars by previous authors \citep{BM87, Fan05}. In this work, we separated the $\gamma$-ray emission into beamed and unbeamed components for 584 {\it Fermi}-LAT blazars. We now discuss some of the implications. 

\subsection{Correlations analysis}

\begin{figure}
   \centering
   \includegraphics[width=8.9cm]{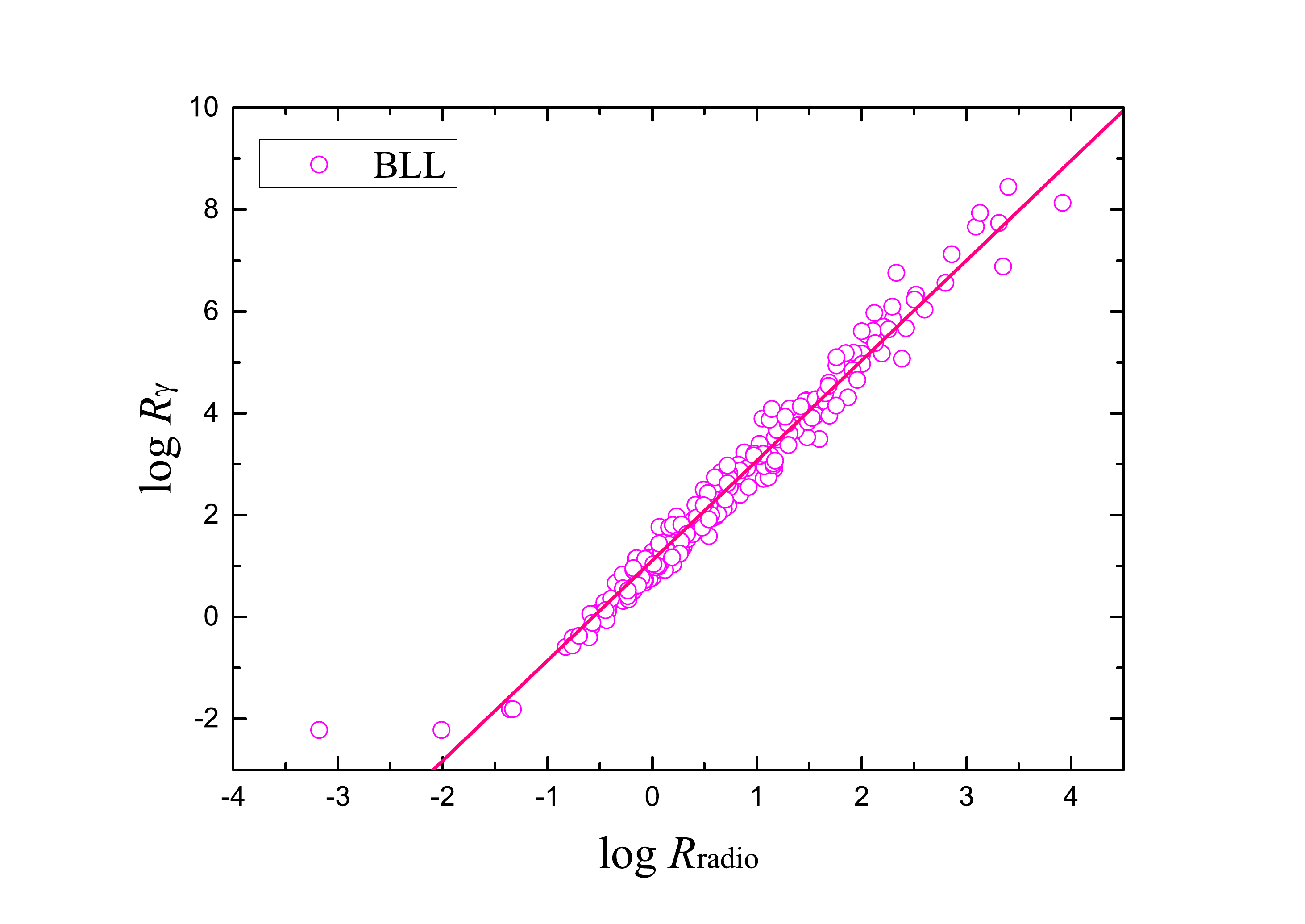}
   \includegraphics[width=8.9cm]{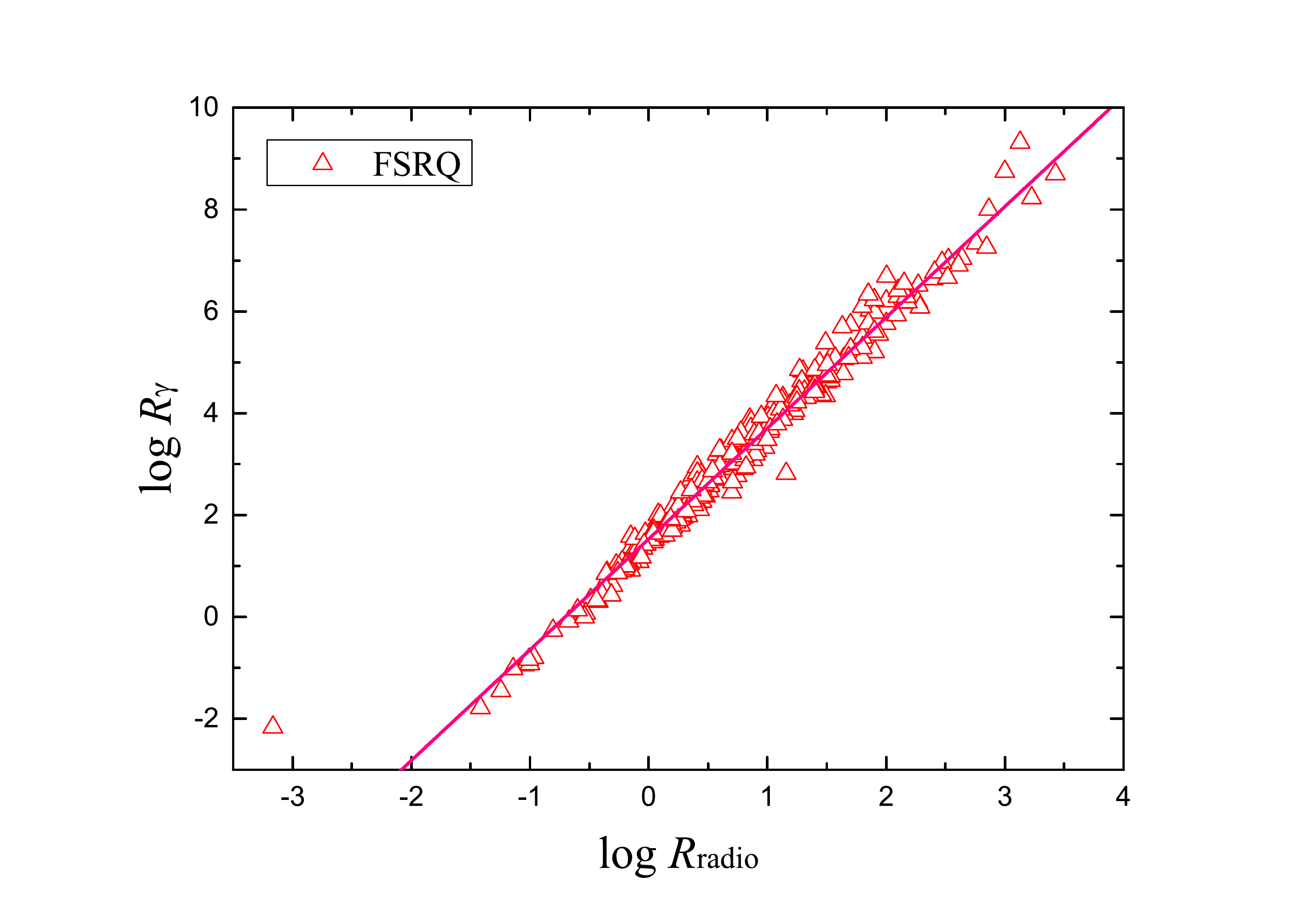}
    \caption{Plot of the $\gamma$-ray core-dominance parameter, $\log R_{\gamma}$, against radio core-dominance parameter, $\log R_{\rm radio}$, for BL Lacs (upper panel) and FSRQs (lower panel). In this Figure, magenta $\bigcirc$ stands for BL Lacs, red $\bigtriangleup$ for FSRQs. The best-fitting indicates that $\log R_{\gamma} = (1.96 \pm 0.02) \log R_{\rm radio} + (1.11 \pm 0.03)$ and $\log R_{\gamma} = (2.18 \pm 0.02) \log R_{\rm radio} + (1.53 \pm 0.02)$ for BL Lacs and FSRQs, respectively.}
    \label{fig4}
\end{figure}

Fig. \ref{fig4} shows the correlation between the $\gamma$-ray core-dominance parameter $\log R_{\gamma}$ and the radio core-dominance parameter $\log R_{\rm radio}$ for our sample. The best-fitting results are $\log R_{\gamma} = (1.96 \pm 0.02) \log R_{\rm radio} + (1.11 \pm 0.03)$ with a correlation coefficient $r= 0.98$ and a chance probability of $p\sim0$ for BL Lacs, and $\log R_{\gamma} = (2.18 \pm 0.02) \log R_{\rm radio} + (1.53 \pm 0.02)$ with $r= 0.99$ and $p\sim0$ for FSRQs. We listed the linear regression results in Table \ref{tab2}. 

By modeling the radio light curves of 1029 sources as a series of flares characterized by an exponential rise and decay, \citet{Lio18} estimated the variability Doppler factor for 837 blazars (167 BL Lacs, 670 FSRQs). There are 285 sources in common in our sample with the work of \citet{Lio18} including 75 BL Lacs and 210 FSRQs. When we consider the relation between $\log R_{\gamma}$ and $\delta$, two positive tendencies are found with correlation coefficient of $r=0.22$ for BL Lacs and $r=0.30$ for FSRQs, respectively. Their chance probabilities are $p<10^{-4}$. These results show that the $\gamma$-ray core-dominance parameter can be a statistical indicator of Doppler boosting.

\begin{table}
\caption{\label{tab2} The linear regression analysis results ($y=(a\pm\Delta a)x+(b\pm\Delta b)$) for correlations between two parameters sample}
\centering
\setlength{\tabcolsep}{5mm}{
\begin{tabular}{cccccc}
\hline\hline
$y$ versus $x$ & Sample & $a\pm\Delta a$ & $b\pm\Delta b$ & $r$ & $p$ \\
\hline

$\log L_{\gamma}\sim\log L_{\rm r, b}$ & Blazar & $0.70\pm0.02$ & $10.77\pm0.53$ & 0.82 & $\sim0$ \\ 
$\log R_{\gamma}\sim\log R$ & BL & $1.96\pm0.02$ & $1.11\pm0.03$ & 0.98 & $\sim0$\\
                                                & FSRQ   & $2.18\pm0.02$ & $1.53\pm0.02$ & 0.99 & $\sim0$\\
%$\alpha^{\rm ph}_{\gamma}\sim\log R_{\gamma}$$^{\star}$ & Blazar & $0.32\pm0.04$ & $1.41\pm0.12$ & 0.21 & $2\times10^{-7}$\\                                               
%$\log L_{\gamma}\sim\log R_{\gamma}$$^{\star}$ & Blazar & $0.88\pm0.04$ & $26.82\pm0.12$ & 0.12 & $0.0046$\\ 
$\log L_{\rm r, unb}\sim\log L_{\gamma, \rm unb}$ & BL & $0.40\pm0.03$ & $14.25\pm0.72$ & 0.68 & $\sim0$ \\  
                                                                                  & FSRQ & $0.39\pm0.02$ & $16.16\pm0.62$ & 0.70 & $\sim0$ \\

\hline
\end{tabular}
}
%\tablecomments{The correlation regression labelled with a superscript $\star$ are performed by the symmetrical ordinary least-squares (OLS) bisector method. The others are used the ordinary linear fitting.}
\end{table} 

\begin{figure}
   \centering
   \includegraphics[width=10cm]{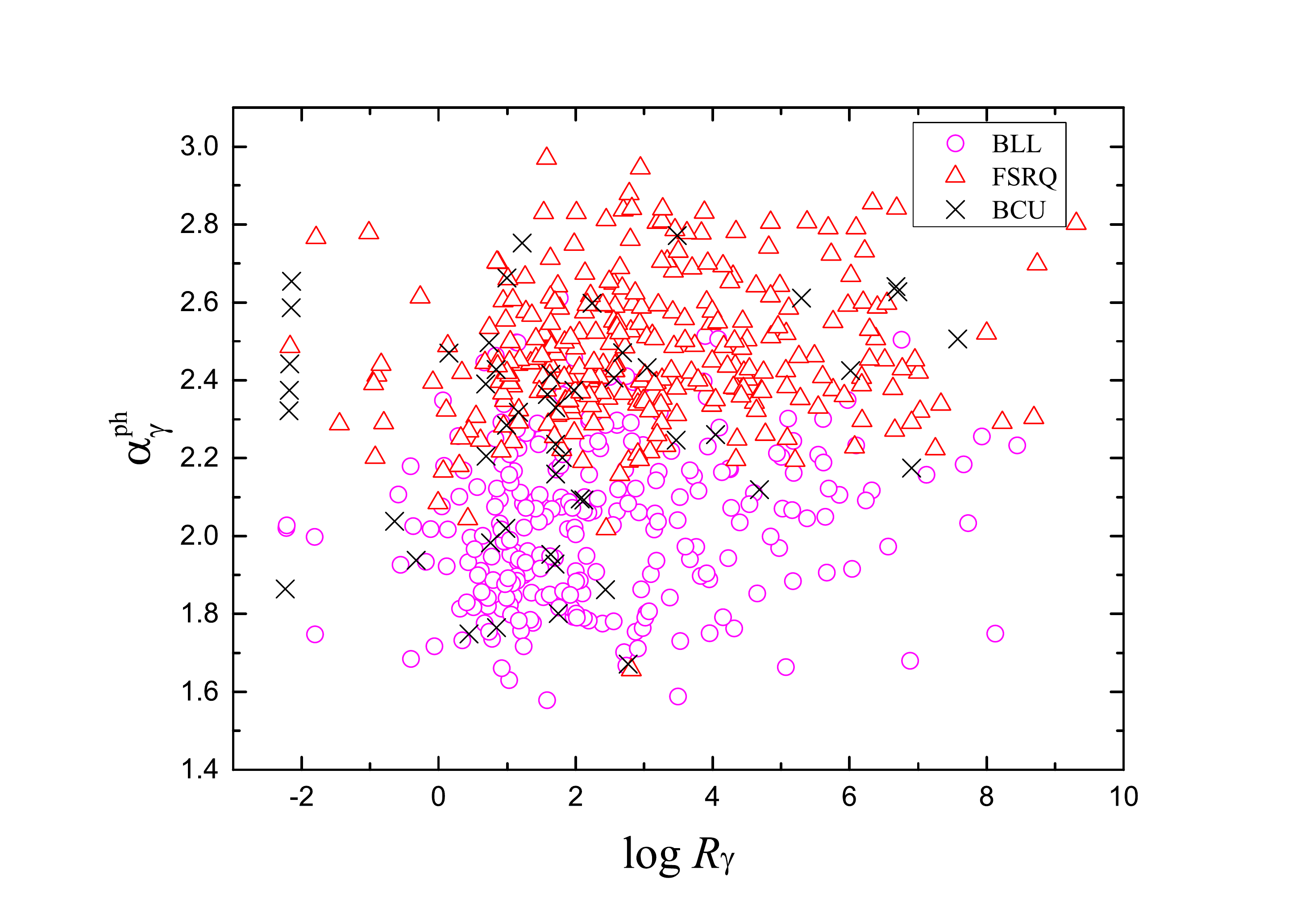}
    \caption{Plot of the $\gamma$-ray photon index, $\alpha^{\rm ph}_{\gamma}$, against $\gamma$-ray core-dominance parameter, $\log R_{\gamma}$, for the whole sample.} %The best-fitting here is performed by the ordinary least-squares (OLS) bisector method and it demonstrates that $\alpha^{\rm ph}_{\gamma} = (0.32 \pm 0.04) \log R_{\gamma} + (1.41 \pm 0.12)$. In this Figure, magenta $\bigcirc$ stands for BL Lacs, red $\bigtriangleup$ for FSRQs and black $\times$ for BCUs.}
      \label{fig5}
\end{figure}

\citet{K93} found that the optical-X-ray spectral index ($\alpha_{\rm ox}$) decreased with the increasing ratio of the beamed X-ray component to the unbeamed one, i.e. X-ray beaming factor, and the correlation was $\alpha_{\rm ox}=-0.023\log\frac{L_{\rm x, b}}{L_{\rm x, unb}}+1.3$, implying that highly beamed radio sources have stronger X-ray emission for a given optical luminosity. We also show in Fig. \ref{fig5} the plot of $\gamma$-ray photon index $\alpha^{\rm ph}_{\gamma}$ against $\log R_{\gamma}$. We found that the scatter is quite large. 
%so that we may not use the ordinary least-squares (OLS) regression method to fit all of points as we usually do, instead, we consider a symmetrical OLS bisector method\footnote{The computer code is available in \url{https://astrostatistics.psu.edu/statcodes/sc_regression.html}}\citep{Iso90, FB92}. When the OLS bisector performs in this plot, we obtain the best-fitting is $\alpha^{\rm ph}_{\gamma} = (0.32 \pm 0.04) \log R_{\gamma} + (1.41 \pm 0.12)$ with $r= 0.21$ and $p = 2 \times 10^{-7}$ (see Table \ref{tab2}). This significant correlation shows a trend for steeper $\gamma$-ray spectra corresponding to higher $\log R_{\gamma}$.    

\begin{figure}
   \centering
   \includegraphics[width=10cm]{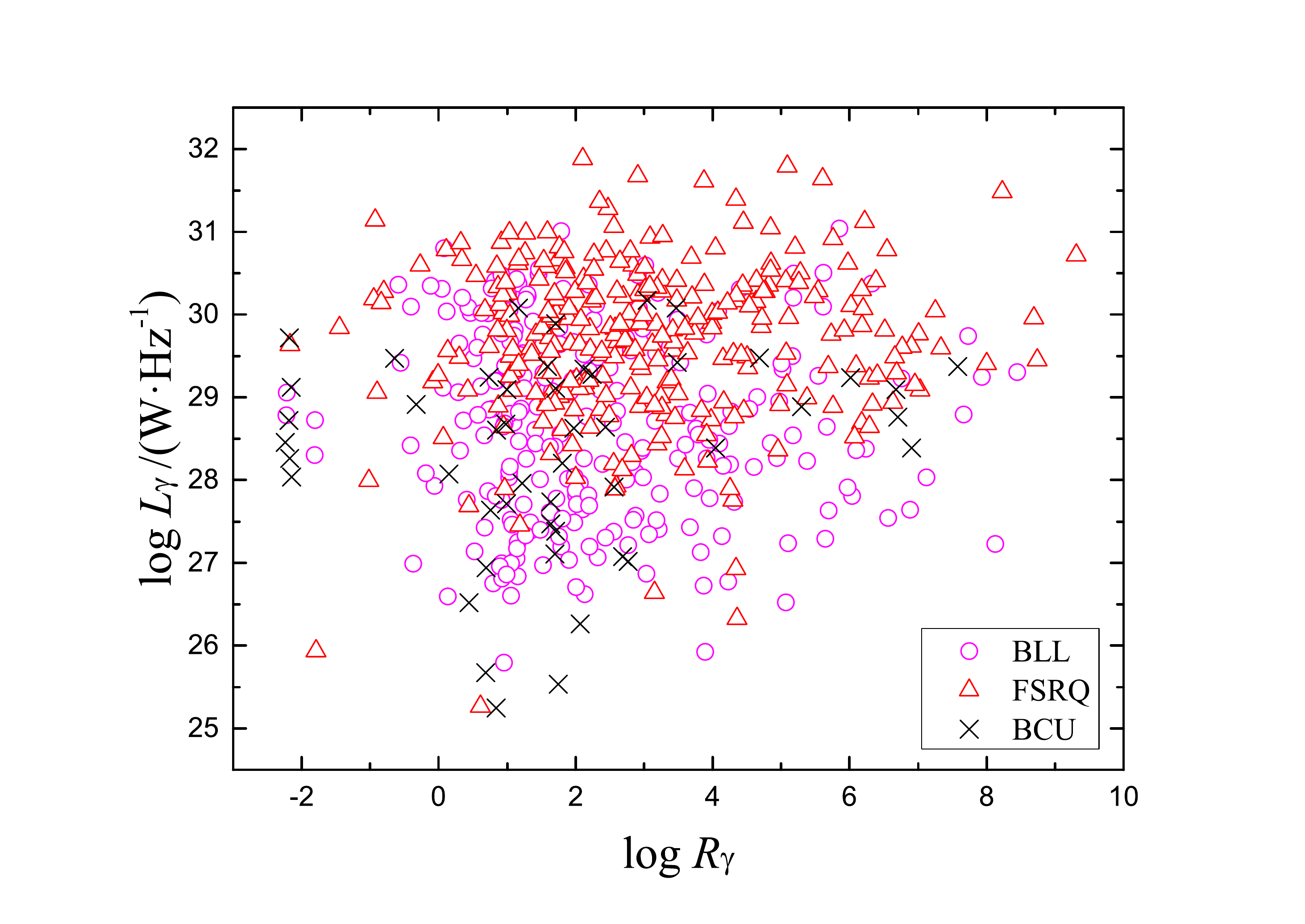}
    \caption{Plot of the $\gamma$-ray luminosity, $\log L_{\gamma}$, against $\log R_{\gamma}$.} %The best-fitting gives $\log L_{\gamma} = (0.88 \pm 0.04) \log R_{\gamma} + (26.82 \pm 0.12)$ when we adopt the OLS bisector.}
      \label{fig6}
\end{figure}

\cite{K93} also discussed the relation between X-ray luminosity and X-ray beaming factor, and obtained $\log L_{X}=0.124\log\frac{L_{X, \rm b}}{L_{X, \rm unb}}+20.71$. They concluded that the X-ray luminosity increased with the ratio of the beamed X-ray component to the unbeamed component (i.e. X-ray beaming factor). Following this idea, we show the plot of the $\gamma$-ray luminosity $\log L_{\gamma}$ against $\log R_{\gamma}$ in Fig. \ref{fig6}.
%Here we also adopt the OLS bisector method due to the high degree of scatter dispersion, and the linear regression leads to $\log L_{\gamma} = (0.88 \pm 0.04) \log R_{\gamma} + (26.82 \pm 0.12)$ with $r= 0.12$ and $p=0.0046$ (see also Table \ref{tab2}). This result predicts that, on average, $\gamma$-ray emission increases tardily from the weakly beamed sources to the highly beamed ones.   

In the two-component model, the total radio luminosity, $L_{\rm tot}$, is separated into the core and the extended components, i.e. $L_{\rm tot} = L_{\rm core} + L_{\rm ext} = R_{\rm radio}L_{\rm ext} + L_{\rm ext} = (1+R_{\rm radio})L_{\rm ext}$, therefore one can expect that $L_{\rm tot} / L_{\rm ext} \propto (1+R_{\rm radio})$. \citet{WuD14} found that $\log (L_{\gamma}/L_{\rm ext}) \sim 0.946 \log (1+R_{\rm radio})$ for 124 $\gamma$-ray sources collected from 2FGL \citep{2FGL}. \citet{Pei20(a)} also obtained that $\log (L_{\gamma}/L_{\rm ext}) \sim 0.92 \log (1+R_{\rm radio})$ for the present {\it Fermi} blazar sample from 4FGL. This supports the idea that $\gamma$-ray emission consists of two-component alike the radio one. 

\begin{figure}
   \centering
   \includegraphics[width=8.9cm]{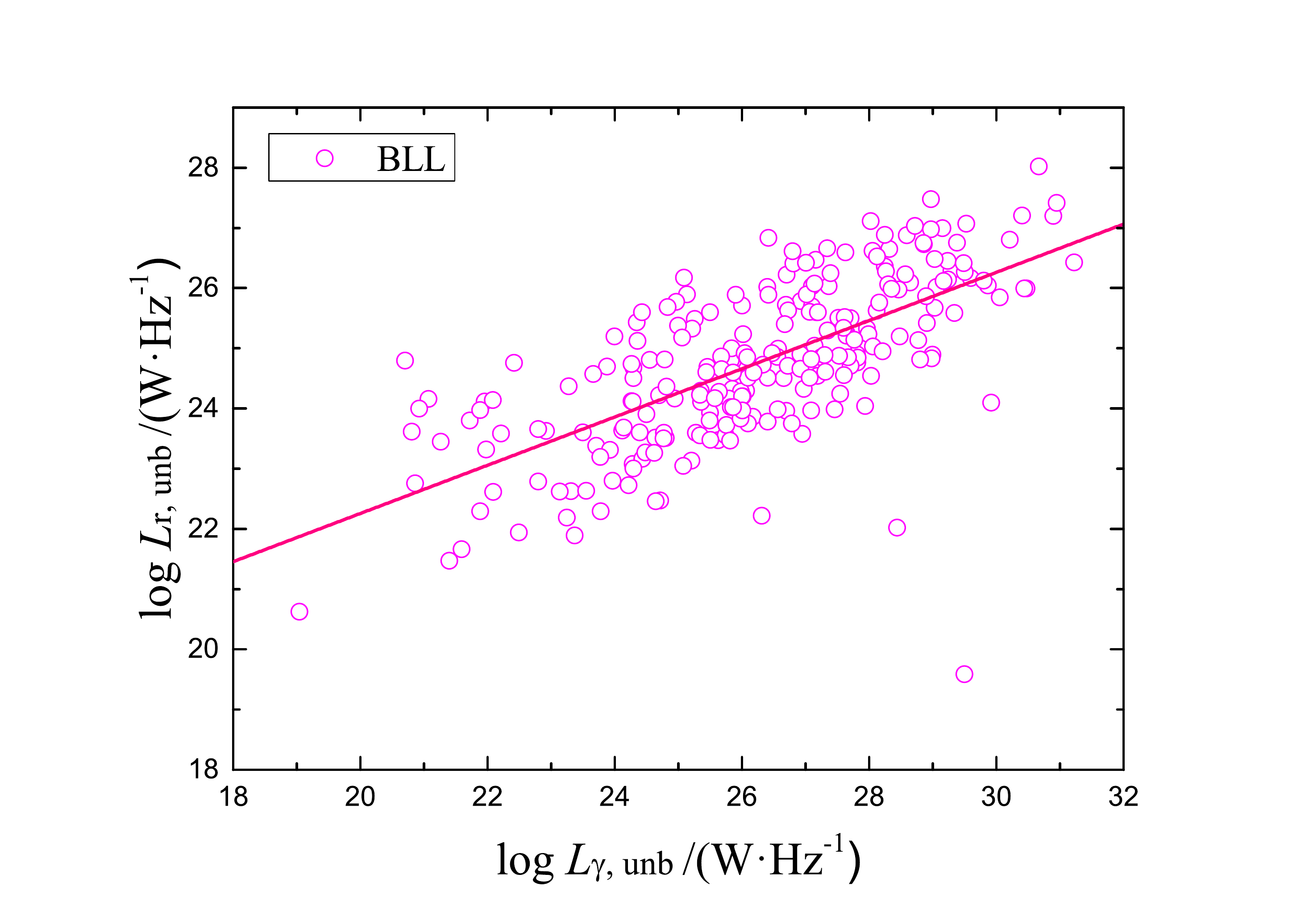}
   \includegraphics[width=8.9cm]{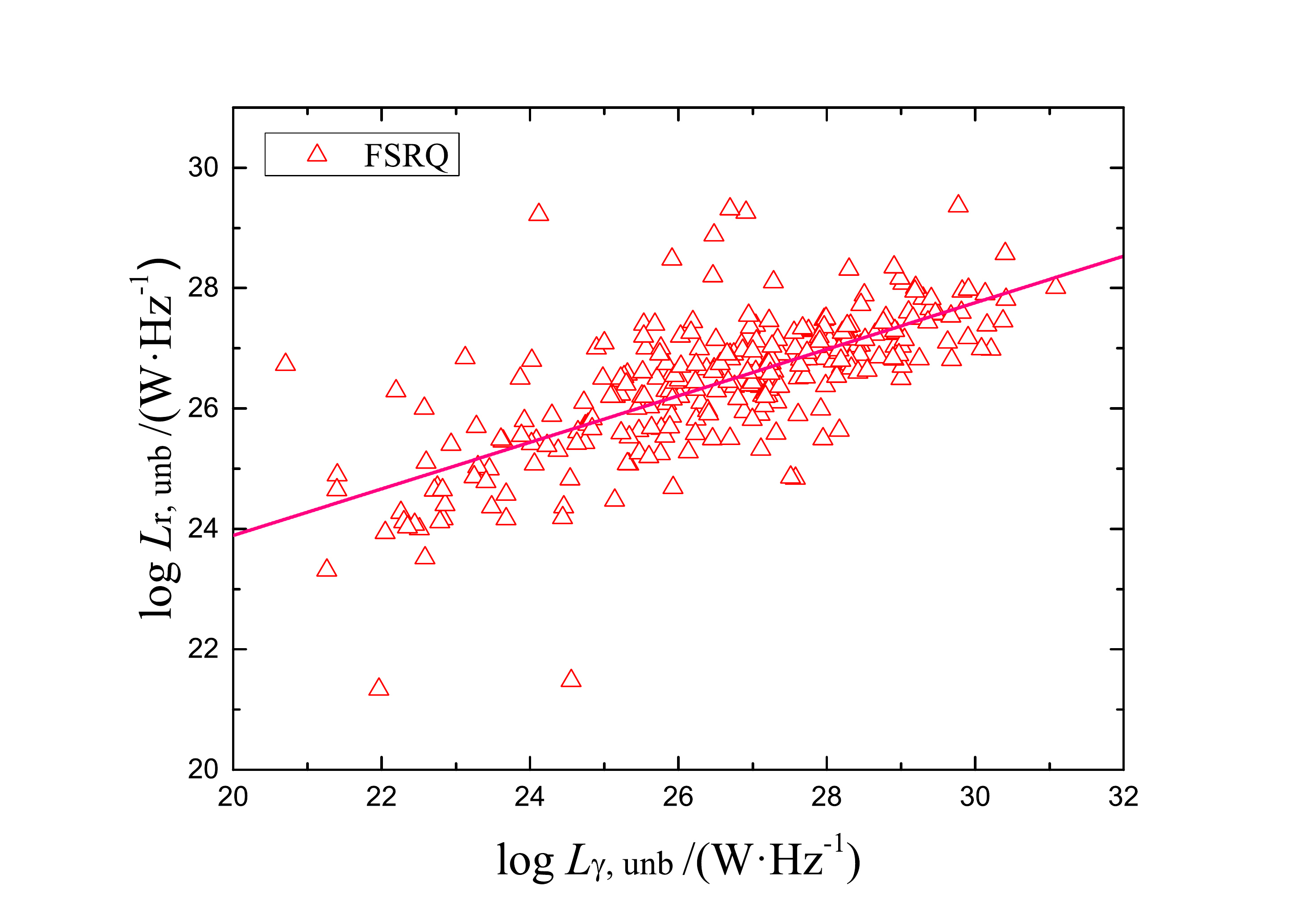}
    \caption{Plot of the extended radio luminosity ($\log L_{\rm r, unb}$), against extended (unbeamed) $\gamma$-ray luminosity ($\log L_{\gamma, \rm unb}$) for BL Lacs and FSRQs. The best-fitting signifies that $\log L_{\rm r, unb} = (0.40 \pm 0.03) \log L_{\gamma, \rm unb} + (14.25 \pm 0.72)$ for BL Lacs and $\log L_{\rm r, unb} = (0.39 \pm 0.02) \log L_{\gamma, \rm unb} + (16.16 \pm 0.62)$ for FSRQs.}
      \label{fig7}
\end{figure}

Consequently, when we apply the two-component model to the $\gamma$-ray emission, i.e.\ $L^{\rm tot}_{\gamma} = L_{\gamma, \rm b} + L_{\gamma, \rm unb}$, and considering the definition of $\gamma$-ray core-dominance parameter, then we have
\begin{equation}
1+R_{\gamma} = 1+\displaystyle \frac{L_{\gamma, \rm b}}{L_{\gamma, \rm unb}}= \displaystyle \frac{L^{\rm tot}_{\gamma}}{L_{\gamma, \rm unb}}, \,\,\,\,
\log L_{\gamma, \rm unb} = \log L^{\rm tot}_{\gamma} - \log (1+R_{\gamma}) 
\label{eq13}
\end{equation} 
Thus we can obtain the extended (unbeamed) $\gamma$-ray luminosity by subtracting $(1+R_{\gamma})$ from the total $\gamma$-ray luminosity. For the blazars in our sample, the unbeamed $\gamma$-ray luminosity, $\log L_{\gamma, \rm unb}$ (W Hz$^{-1}$) is found to be correlated with the extended radio luminosity, $\log L_{\rm r, unb}$ (W Hz$^{-1}$) as shown in Fig. \ref{fig7}. The best-fittings are $\log L_{\rm r, unb} = (0.40 \pm 0.03) \log L_{\gamma, \rm unb} + (14.25 \pm 0.72)$ with $r= 0.68$ and $p \sim 0$ for BL Lacs, and $\log L_{\rm r, unb} = (0.39 \pm 0.02) \log L_{\gamma, \rm unb} + (16.16 \pm 0.62)$ with $r= 0.70$ and $p \sim 0$ for FSRQs. However, these correlations have a redshift dependence. Thus the redshift effect needs to be removed. To do so, we adopt a partial correlation analysis \citep[see e.g.][]{Pad92}, 
\begin{equation}
r_{12,3} = \displaystyle\frac{r_{12}-r_{13}r_{23}}{\sqrt{1-r^{2}_{13}}\sqrt{1-r^{2}_{23}}}
\label{eq14}
\end{equation} 
where $r_{ij}$ denotes the correlation coefficient between $x_{i}$ and $x_{j}$, whilst $r_{ij,k}$ denotes the partial correlation coefficient between $x_{i}$ and $x_{j}$ with $x_{k}$ dependence excluded ($i, j, k=1,2,3$). In our case, we let $x_{1}=\log L_{\rm r, unb}$, $x_{2}=\log L_{\gamma, \rm unb}$ and $x_{3}=z$. For FSRQs, we have $r_{12}=0.70$, $r_{1z}=0.50$ and $r_{2z}=0.24$, which yields $r_{12,z}= 0.69$. Using the similar calculation, we obtain $r_{12,z}= 0.66$ for BL Lacs. The $p$-value is $\sim0$ in both cases. It still shows a statistically significant correlation between $\log L_{\rm r, unb}$ and $\log L_{\gamma, \rm unb}$ after removing the redshift effect, suggesting that the extended emission in radio and $\gamma$-rays are truly correlated.

\subsection{Differences in Doppler boosting between FSRQs and BL Lacs}

The radio core-dominance parameter is associated with Doppler factor as in Fig. \ref{fig1}. One can anticipate that a source with larger Doppler factor should have a higher radio core-dominance parameter \citep{Hov09}. Previous studies showed that FSRQs are more Doppler boosted with regard to BL Lacs as evidenced by their larger parsec-scale apparent jet speeds ($\beta_{\rm app}$, defined by $\beta_{\rm app}=\beta\sin\phi/(1-\beta\cos\phi)$) \citep[e.g.,][]{KLC10}. In particular, {\it Fermi}-LAT-detected FSRQs have faster apparent jet speeds \citep{Lis09} and higher Very Long Baseline Array (VLBA) core brightness temperatures than those FSRQs that are not detected by {\it Fermi}-LAT \citep[e.g.][]{Kov09}. 

Based on the study of the kiloparsec-scale radio emission properties of 135 radio-loud AGNs in the MOJAVE\footnote{Monitoring Of Jets in Active galactic nuclei with VLBA Experiments: \url{http://www.physics.purdue.edu/astro/MOJAVE/}} sample of blazars, \citet{KLC10} pointed out that the parsec-scale apparent jet speed and the kiloparsec-scale radio core luminosity are related by the standard beaming relation $L=L_{\rm int}\times\delta^{\lambda}$, where $L_{\rm int}$ is the intrinsic luminosity. The best-fitting values for the sample gave $\delta\simeq\Gamma=52$ (since $\theta$ is very small in the circumstances), $L_{\rm int}=5\times10^{24}$ and $\lambda\approx2$. They found that most quasars have faster jets than most BL Lac in consideration of a large overlap in radio powers between quasars and BL Lacs \citep[see fig. 5 in][]{KLC10}. Additionally, they also found that, after removing the effects of luminosity distance and radio core luminosity, the parsec-scale apparent jet speeds are still correlated tightly with the extended radio luminosity. This implies that more radio powerful sources have faster radio jets and also reveals the fact that faster jets are launched in AGNs with larger kiloparsec-scale lobe luminosities.

\section{CONCLUSION}     \label{sec5}

The {\it Fermi}-Large Area Telescope ({\it Fermi}-LAT) has revolutionized our view of the $\gamma$-ray sky. They are collated into the latest {\it Fermi} catalog, 4FGL, which includes 5065 sources based on the first 8 years of data. AGNs are the vast majority of the catalog entries and $98\%$ of AGNs are blazars \citep{4FGL,4LAC}.        

Standard beaming models expect that more core-dominated sources should be more beamed and have larger Doppler boosting factors. In this work, we have separated the beamed and unbeamed contributions to the total $\gamma$-ray emission for a sample of 584 {\it Fermi}-detected blazars with available radio core-dominance parameters \citep{Pei20(a)}, suggesting that the two-component model could be successfully adopted for the $\gamma$-ray emission of blazars. The main conclusions of this work are the following:

\begin{enumerate}
\item We calculated the $\gamma$-ray core-dominance parameter ($R_{\gamma}$), separating the beamed and unbeamed contributions of $\gamma$-ray emission for 584 {\it Fermi}-LAT blazars, obtaining $\langle \log R_{\gamma} \rangle|_{\mathrm{BL\,Lac}} = 2.38 \pm 1.91$, $\langle \log R_{\gamma} \rangle|_{\rm FSRQ} = 3.07 \pm 2.01$. 

\item The $\gamma$-ray core-dominance parameter $\log R_{\gamma}$ is correlated tightly with the radio core-dominance parameter $\log R_{\rm radio}$, indicating that, the core-dominance parameter is a good statistical indicator of the beaming effect.

\item The unbeamed $\gamma$-ray emission and extended radio emission are also correlated.

\item We conclude that the $\gamma$-ray emission is mainly from the core (or the beamed component).
\end{enumerate}

\begin{acknowledgements}
We thank the anonymous referee for valuable comments and suggestions, which help us to improve the manuscript. We also greatly appreciated the referee for carefully correcting the words and sentences.  
Author Z. Y. Pei acknowledges the ongoing support from Guangzhou University, China, and the University of Padova, Italy. Thanks are given to Prof. Gaoyong Luo from Buckinghamshire New University for the contribution on improving the language of our manuscript. This work is partially supported by the National Natural Science Foundation of China (NSFC 11733001, NSFC U203120008, NSFC U1531245), Natural Science Foundation of Guangdong Province (2017A030313011; 2019B030302001), supports for Astrophysics Key Subjects of Guangdong Province and Guangzhou City.

\end{acknowledgements}

\bibliography{Pei}{}
\bibliographystyle{aasjournal}

\end{document}